\documentclass[aps,prb,reprint,twocolumn, superscriptaddress]{revtex4-1}
\usepackage{bbold}
\usepackage{amsmath,amsfonts,amsmath,mathtools,bbm,bm}
\usepackage{graphicx}
\usepackage{braket}
\usepackage[colorlinks,linkcolor=blue,urlcolor=blue,citecolor=blue]{hyperref}
\usepackage[all]{hypcap}

\usepackage{blindtext}
\usepackage{enumitem}
\usepackage{xcolor}
\usepackage[normalem]{ulem}
\usepackage{float}
\setcitestyle{numbers,square}

\newcommand{\ii}{\mathrm{i}}
\newcommand{\dd}{\mathrm{d}}
\newcommand{\Tr}{\mathrm{Tr}}

\newcommand{\aveE}[1]{\langle\!\langle #1\rangle\!\rangle }

\def\ii{\mathrm{i}}

\def\({\left (}
\def\){\right )}

\graphicspath{{Figure/PNG/}{Figure/PDF/}{Figure/EPS/}{Figure/TEX/}{Figure/}}

\newcommand\mydots{\hbox to 1em{.\hss.\hss.}}

\begin{document}

\title{Microcanonical Free Cumulants in lattice systems}

\author{Felix Fritzsch}
\affiliation{Physics Department, Faculty of Mathematics and Physics, University of Ljubljana,  SI-1000, Slovenia}
\affiliation{Max Planck Institute for the Physics of Complex Systems, 01087 Dresden, Germany}

\author{Toma\v{z} Prosen}
\affiliation{Physics Department, Faculty of Mathematics and Physics, University of Ljubljana,  SI-1000, Slovenia}
\affiliation{Institute of Mathematics, Physics and Mechanics, Ljubljana,  SI-1000, Slovenia}

\author{Silvia Pappalardi}
\email{pappalardi@thp.uni-koeln.de}
\affiliation{Institut f\"ur Theoretische Physik, Universit\"at zu K\"oln, Z\"ulpicher Straße 77, 50937 K\"oln, Germany}

\date{\today}

\begin{abstract}
Recently, the full version of the Eigenstate Thermalization Hypothesis (ETH) has been systematized using Free Probability. In this paper, we present a detailed discussion of the Free Cumulants approach to many-body dynamics within the microcanonical ensemble. Differences between the later and canonical averages are known to manifest in the time-dependent fluctuations of extensive operators. Thus, the microcanonical ensemble is essential to extend the application of Free Probability to the broad class of extensive observables. We numerically demonstrate the validity of our approach in a non-integrable spin chain Hamiltonian for extensive observables at finite energy density. Our results confirm the full ETH properties, specifically the suppression of crossing contributions and the factorization of non-crossing ones, thus demonstrating that the microcanonical free cumulants encode ETH smooth correlations for both local and extensive observables.
\end{abstract}

\maketitle

\section{Introduction}

The Eigenstate Thermalization Hypothesis (ETH) is the acknowledged framework for understanding thermalization and dynamics in many-body nonintegrable systems \cite{deutsch1991quantum, srednicki1999approach, dalessio2016from}.
ETH conjectures matrix elements of physical local observables in the energy eigenbasis possess statistical properties with \emph{smooth dependence} on the involved energies. 
This smoothness allows to substitue canonical averages of one-point and two-time local observables  with the smooth functions encoded by single ETH eigenstates, as demonstrated by extensive numerical evidence 
\cite{prosen1999,rigol2008thermalization, biroli2010effect,  polkovnikov2011colloquium, ikeda2013finite, steinigeweg2013eigenstate,alba2015eigenstate, beugeling2015off, luitz2016long, luitz2016anomalous, leblond2020eigenstate, brenes2020eigenstate, fritzsch2021eigenstate, garratt2021pairing}.

Recently, there has been growing attention on multi-time correlation functions, which encode, e.g., the scrambling of quantum information in terms of out-of-time ordered correlators \cite{maldacena2016bound, hosur2016chaos, roberts2017chaos, xu2022scrambling, brenes2021out, garcia2022out}, higher-order hydrodynamics beyond the linear response regime \cite{doyon2020fluctuations, myers2020transport, fava2021hydrodynamic}, as well as deep thermalization \cite{cotler2023emergent,ho2022exact, claeys2022emergent, ippoliti2022solvable, lucas2022generalized}.
To systematically characterize such multi-time correlation function, a full version of the ETH ansatz has been introduced \cite{foini2019eigenstate}, which captures a hierarchy of correlations between matrix elements of physical observables within energy eigenstates not present in standard ETH.
The combinatorics involved in the full version of ETH ansatz naturally connect it with the mathematical field of Free Probability (FP) of non-commuting random variables, e.g., random matrices \cite{speicher1997free}.
More specifically, free cumulants, a recursively defined generalization of classical cumulants to the setting of Free Probability, appear as the elementary building blocks of multi-time correlation functions \cite{pappalardi2022eigenstate}.
The decomposition of multi-time correlation functions into free cumulants is facilitated by full ETH, which provides a simple, non-recursive representation of free cumulants.
This can be established by demonstrating that the ETH result coincides with the free cumulants at the lowest order and obeys the same recursion relation. Moreover, the structure of free cumulants seems applicable in many contexts in many-body quantum dynamics \cite{cipolloni2022thermalisation, fava2023designs, jindal2024generalized, ampelogiannis2024clustering}.

In an accompanying paper \cite{ourLetter}, we have demonstrated numerically the decomposition of canonical multi-time correlation functions predicted by ETH and Free Probability in chaotic lattice systems, provided that the observable is local (supported on a few adjacent sites). 
For local observables, the expectation values in single energy eigenstates are expected to coincide with the respective averages within equilibrium ensembles, including the canonical and microcanonical ensemble as well as the diagonal ensemble following a quantum quench \cite{rigol2008thermalization, essler2016quench}. \\
Often, however, one is interested in
situations where the observables have an extensive nature.
In this situation, and more generally, the eigenstate-to-eigenstate fluctuations, as well as the energy fluctuations, lead to deviations between ensemble averages and single eigenstate results.
Those deviations are usually small on the level of one-point functions, i.e., expectation values of physical observables, but they become relevant already for two-time connected correlations of extensive observables in the canonical ensemble  \cite{dalessio2016from}. To address this issue, one shall consider averages on the microcanonical shell, where the energy fluctuations can be made arbitrarily small.

In this work, we generalize the results from Refs.\cite{pappalardi2022eigenstate, ourLetter} to the \emph{microcanonical ensemble} and corroborate it by numerical demonstrations in the case of extensive observables. 
The study of many-body dynamics in microcanonical energy shells has recently gauged renewed attention via the use of so-called filtering techniques to evaluate expectation values both conceptually \cite{richter2020eigenstate, pappalardi2024microcanonical} and numerically \cite{long2003finite, zotos2006microcanonical, okamoto2018accuracy, sirui2021algorithms, yilun2022classical}. 
Here, we carefully study the microcanonical scenario and identify entropic contributions to the energy fluctuation as the main reason for the failure of both full and standard ETH in the canonical ensemble.
We contrast this with full ETH in the microcanonical ensemble,
in which the freedom of choosing the width of the microcanonical energy shell allows for suppressing energy fluctuations.
We demonstrate that this eventually restores the validity of both standard and full ETH, and we provide a detailed description of microcanonical free cumulants and the corresponding decomposition of out-of-time-ordered four-time correlation functions for extensive observables.\\

The rest of the paper is organized as follows. 
In Sec.~\ref{sec_fluctua} we contrast standard ETH in the canonical ensemble with the microcanonical ensemble.
Subsequently, we provide an extensive discussion of full ETH in the microcanonical ensemble in Sec.~\ref{sec:fullETH} and test its predictions numerically in Sec.~\ref{sec_num}. We eventually conclude in Sec.~\ref{sec:conclusion}.

\section{Canonical and microcanonical fluctuations in the standard ETH}
\label{sec_fluctua}

We consider a non-integrable Hamiltonian $\hat H$ of $N$ locally coupled degrees of freedom with spectrum $\hat H\ket{E_i} = E_i \ket{E_i}$.
In this paper, we study averages over equilibrium ensembles at energy $E$ denoted by
\begin{equation}
	\langle \,  \cdot \, \rangle_E \equiv \text{Tr}(\hat \rho_E \, \cdot \, )\ ,
 \label{eq:ensemble_average}
\end{equation}
where $ \hat \rho_E $ is an equilibrium density matrix, diagonal in energy, of the form 
\begin{equation}
    \label{rho_E}
    \hat \rho_E = \sum_{i} \frac{p_{E}(E_i)}{Z_E} \ket{E_i}\bra{E_i}\ ,
\end{equation}
 where $p_{E}(E_i)$ is a generic smooth function that strongly peaks around some energy $E$. The normalization is $Z_E=\sum_i p_{E}(E_i)$ such that $\langle \mathbb{1} \rangle_E = 1$. This naturally includes the standard equilibrium ensembles, i.e., the canonical and microcanonical ensembles, as well as the diagonal ensemble after a quantum quench \cite{rigol2008thermalization, essler2016quench}.

We will compare such ensemble averages $\langle \cdot \rangle_E$ in Eq.~\eqref{eq:ensemble_average} with expectation values with respect to single eigenstates $\langle E_i|\cdot \ket{E_i}$ of systems obeying the Eigenstate Thermalization Hypothesis \cite{srednicki1994chaos, dalessio2016from}.
The latter conjectures matrix elements $A_{ij}=\bra{E_i}\hat A\ket{E_j}$ of physical observables $\hat A$, e.g. local observables or extensive sums thereof, in the eigenbasis $\ket{E_i}$ of a Hamiltonian $\hat H$ to smoothly depend on the energies $E_i$.
In its standard formulation, the ETH reads
\begin{equation}
    \overline{A_{ii}} = F^{(1)}_{e^+} = \mathcal{A}(e^+) \quad \text{and} \quad \overline{A_{ij}A_{ji}} = e^{-S(E^+)}F^{(2)}_{e^+}(\omega_{ij}),
    \label{eq:ETH2}
\end{equation}
for $i\neq j$ and $\overline{A_{ii}A_{ii}} = \overline{A_{ii}}^2$ at the leading order in $\mathcal O( e^{-S(E^+)})$.
Here, $E^+ = (E_i + E_j)/2$ is the average energy, corresponding to energy density $e^+=E^+/N$ and $\omega_{ij}=E_i - E_j$ is the associated frequency.
Moreover, $S(E^+)=Ns(e^+)$ is the extensive thermodynamic entropy, $F^{(1)}_{e^+} = \mathcal{A}(e^+)$ ($F^{(2)}_{e^+}(\omega_{ij})$) is a smooth function of energy density (and frequency).
The bar indicates some average over a fictitious ensemble, e.g., over a small energy window or an ensemble of systems with similar physical properties \cite{deutsch1991quantum, foini2019eigenstate}.
This average shall be intended as a statement about the self-averaging properties of the matrix elements and is not to be confused with the average over equilibrium ensembles we introduce next. \\

In this introductory section, we will review how the standard canonical and the microcanonical averages may lead to different results at the level of thermal two-time correlation functions of extensive observables.

\subsection{Averages over smooth densities}
\label{sec:scaling}


We indicate with $F(E_i)$ a generic smooth function of an eigenstate $E_i$ (for instance, for diagonal matrix elements $A_{ii}$ as well as their fluctuations $ A_{ii}^2$ or $A_{ij}  A_{ji}$).
In a system obeying ETH, these quantities are smooth functions $\mathcal F(e_i)$ of the energy densities $e_i=E_i/N$ (for instance $\mathcal A(e_i)^2$ or $F^{(2)}_{e_i-\omega/2N}(\omega)$ via Eq.~\eqref{eq:ETH2})
but may fluctuate from eigenstate to eigenstate. 
The corresponding ensemble average~\eqref{eq:ensemble_average} can be written as a \emph{smoothed average} 
 \begin{equation}
     \aveE{F(E_i)}=\sum_i F(E_i)  \frac{p_{E}(E_i)}{Z_E} \ ,
     \label{eq:smoothed_average}
 \end{equation}
 where $p_{E}(E_i)/Z_E$ is the smooth distribution defined in Eq.~\eqref{rho_E}. We emphasize that this average, taken with respect to the ensemble defined by $p_E(E_i)$ and indicated by the subscript $E$, does not coincide with the average statistical properties of matrix elements in Eq.~\eqref{eq:ETH2}.
Using the smoothness of $F(E_i)$, one can substitute it under summation with $\mathcal F(e_i)$.  In the thermodynamic limit, it is customary to evaluate the smoothed average, Eq.~\eqref{eq:smoothed_average}, by replacing the sum with integrals as
 \begin{equation}
     \frac{1}{Z_E}\sum_i \mathcal F(E_i)p_E(E_i) = \frac{1}{Z_E}\int \dd E^\prime e^{S(E^\prime)} p_E(E^\prime) \mathcal F(E^\prime)
\label{eq:smoothed_averages_integral}    
\end{equation}
and by solving the resulting integral by means of the stationary phase approximation.

Since the ensembles are assumed to be strongly peaked around some extensive energy $E=Ne$, one 
can Taylor expand $\mathcal F(E_i)$ around $E_i\simeq Ne$ and obtain that the smoothed average is dominated by the value of the function evaluated at the energy density $e$ as
\begin{align}
        \begin{split}
    \label{eq_Oe}
    \aveE{F(E_i)}
     & = \mathcal F(e) + \Delta_E^{(1)} \mathcal F'(e) + \frac 12 \Delta_E^{(2)} \mathcal F''(e) + \dots \, .        
    \end{split}
\end{align}
Here,  $\mathcal F'(e)=\frac{\partial \mathcal F}{\partial e}(e)$ and $\mathcal F''(e)=\frac{\partial^2 \mathcal F}{\partial e^2}(e)$ denote the first and second derivative with respect to the energy density.
Additionally, $\Delta_E^{(n)}$ denotes the energy fluctuations expressed as the moments of energy density given by
\begin{equation}
    \Delta_E^{(n)} = \aveE{\left(E_i/N - e\right)^n}=
    \langle (\hat H/N -e)^n\rangle_E, \label{eq:energy_fluctuations}
\end{equation}
which controls the deviations between the single-eigenstate and the averaged result.
Summarizing, \emph{averages over equilibrium densities at energy $E$ density, are dominated by the value at that energy, with corrections coming from the energy fluctuations.}

%


\subsection{Canonical and microcanonical averages}

While the discussion applies to arbitrary equilibrium ensembles, we now specialize in the canonical and microcanonical ones to illustrate the differences in energy fluctuations and their scaling behavior with system size $N$.
We start with the \emph{canonical ensemble} at inverse temperature $\beta$ characterized by 
\begin{equation}
	p_{E_\beta}(E_i)= e^{-\beta E_i}
\end{equation}
which possess extensive thermal energy $E_\beta=e_\beta N$.
The energy fluctuations, Eq.~\eqref{eq:energy_fluctuations}, can 
be directly obtained from evaluating Eq.~\eqref{eq:smoothed_averages_integral} for $\mathcal F(E^\prime)=(E^\prime/N - e_\beta )^n$ in stationary phase approximation.
This yields
\begin{equation}
	\label{canonicalFluc}
	\Delta^{(1)}_{E_\beta}=0 \ , \quad 
	\Delta^{(2)}_{E_\beta}= \frac 1{N^2|S''|} \sim \frac 1N
	\ ,
\end{equation}
where one uses that the entropy's second derivative ${S'' =\frac{\partial^2 S}{\partial E^2}=-|S''|}$ is negative due to the convexity of the thermodynamic entropy as well as  $|S''|\sim 1/N$  due to extensivity of $S$ and $E$, see App.\ref{app_Efluc}.
Hence, the canonical ensemble's distribution is well peaked around energy $E_\beta$ in the large $N$ limit, with small reduced fluctuations.
As a consequence, smoothed averages within the canonical ensemble differ from $\mathcal F(E_\beta)$ by 
\begin{equation}
    \label{cano_corre}
    \langle\!\langle F(E_i)\rangle\!\rangle_{E_\beta}
   = \mathcal F(e_\beta) + \mathcal O(1/N) \ .
\end{equation}
As we review below, the correction $\mathcal O(1/N)$ may become relevant in the computation of connected correlations of extensive observables \cite{dalessio2016from}.  \\

The result is different in the \emph{microcanonical ensemble} at energy $E$ with width $\Delta$.
For definiteness, we consider a Gaussian smoothed window
\begin{equation}
\label{DOS}
p_{E, \Delta}(E_i) 
= e^{-\frac{(E_i-E)^2}{2\Delta^2}}
\end{equation}
with $Z_E = \sum_i  e^{-\frac{(E_i-E)^2}{2\Delta^2}}$.
While this choice is convenient for analytical computations, the results do not depend on the particular smoothing of the microcanonical energy shell.
Repeating the calculation for the energy fluctuations yields in stationary phase approximation, see e.g. App.\ref{app_Efluc}, one has
\begin{subequations}
\label{Delta_}
	\begin{align}
		\Delta^{(1)}_E = N S' \Delta^{(2)}_{E} \\
		 \Delta^{(2)}_{E} = \frac 1{N^2}\frac {\Delta^2}{1+\Delta^2|S''|} \ .
	\end{align}
\end{subequations}
Here, the first derivative of the entropy $S'=\frac {\partial S}{\partial E}=\beta_E$ acts as the microcanonical temperature.  The notable feature of the microcanonical ensemble, which is in stark contrast with the canonical ensemble, is that one can take the energy width to be the smallest scale as long as the microcanonical energy shell hosts sufficiently many states to facilitate meaningful averages.
Apart from that constraint, we choose 
\begin{equation}
    \Delta^2 \ll \frac 1{|S''|}\sim N
\end{equation}
which implies that we can neglect $|S''|$ in Eq.~\eqref{Delta_}, resulting in $$\Delta^{(2)}_{E} = \frac {\Delta^2}{N^2}.$$
In other words, in thermodynamically large systems, one can always send $\Delta$ to zero independently of $N$. 
This implies that the smoothed average, Eq.~\eqref{eq_Oe}, is always dominated by $\mathcal F(e)$ with only small corrections, i.e., 
\begin{equation}
    \label{micro_ave}
    \aveE{F(E_i)}_E
   = \mathcal F(e) + \mathcal O(\Delta^2/N) \ .
\end{equation}

Summarizing, \emph{the microcanonical average overcomes entropic contributions and always sets the arguments of the energy averages to $E$}.

\subsection{Fluctuations in the standard ETH}
\label{sec:scaling_fluc}

We will now discuss how the possible differences between ensemble averages, due to the energy fluctuations, affect two-time correlation functions 
in the context of standard ETH, Eq.~\eqref{eq:ETH2}.
To this end, we denote the average of an observable $\hat A$ on a single energy eigenstate $\ket E_i$ as
\begin{subequations}
\label{k1k2_singleE}
\begin{equation}
    k_{1}^{i}  = \bra{E_i} \hat A\ket{E_i}  = A_{ii}
\end{equation}
and the two-time correlations by
\begin{align}
\begin{split}
\label{k2_singleE}
k_{2}^{i}(t) & = \bra{E_i} \hat A(t)\hat A\ket{E_i} - \bra{E_i} \hat A\ket{E_i}^2 
\\ &  = \sum_{j: j\neq i} A_{ij} A_{ji} e^{i (E_i-E_j) t} \ ,
\end{split}
\end{align}  
\end{subequations}
where  
$\hat A(t) = e^{i \hat H t} \hat A e^{-i \hat H t}$ is the time-evolved operator in the Heisenberg picture (we set $\hbar=1$).

The corresponding one point equilibrium average is
\begin{subequations}
\label{k1k2def}
\begin{align}
    \kappa^E_1 & \equiv \langle \hat A \rangle_E =  \aveE{k_1^{} }_E\ ,
\end{align}
that coincides with the smoothed average. This is not the case for the 
 two-time correlation functions, defined as 
\begin{align}
    \kappa^E_2(t)  & \equiv \langle \hat A(t) \hat A\rangle_E 
    - \langle \hat A \rangle_E ^2 
      = \aveE{k_{2}^{{}}(t)}_E + 
     \aveE{A_{ii}^2}_E - \aveE{A_{ii}\, }_E^2  \ .
     \label{k2ethii}
     \end{align}
\end{subequations}
which differs from the smoothed average of the ETH value due to the variance of diagonal matrix elements, i.e. $\aveE{A_{ii}^2}_E - \aveE{A_{ii}\, }_E^2$. This can be estimated from the smoothness of the ETH ansatz.
Applying Eq.~\eqref{eq_Oe} to $\aveE{A_{ii}^2}_E$ and $\aveE{A_{ii}\, }_E^2$ one has
\begin{align}
\begin{split}
    \label{facto2}
	 \aveE{A_{ii}^2}_E
	 & =  \aveE{A_{ii}\, }_E^2 + \mathcal A'(e_i)^2 \Delta^{(2)}_E +\dots \ .
\end{split}
\end{align}
which is a \emph{factorization of smooth averages with repeated indices at the leading order in $\Delta^{(2)}$}, as we will see also in Sec.~\ref{sec:fullETH} below.
Altogether, this shows that the ensemble averaged two-point function differs from the smoothed average of ETH by
\begin{align}
    \label{k2eth}
    \kappa^E_2(t)  & = \aveE{k_{2}^{{}}(t)}_E 
    + \mathcal O(\Delta_E^{(2)})\ .
     \end{align}

 Summarizing, \emph{while the expectation values of any observable in a smooth diagonal ensemble correspond to the average from single eigenstates, fluctuations have essentially two contributions: the first coming from fluctuations within each eigenstate and the second from the energy fluctuations of the ensemble}.

\subsection{Local vs extensive observables}
The contributions that arise from the energy fluctuations in Eqs.\eqref{k2ethii} not only depend on the equilibrium ensemble but on the nature of the observable too. In particular, for certain combinations of observable and ensemble, the corrections might be of the same order as the smooth average $\aveE{k_{2}^{}(t)}_E$ and hence can not be neglected.

In the case of \emph{local or intensive} observables which are confined to a finite subsystem (such as observables $A_{\rm loc}=a_r$ supported on a few  adjacent sites $r$), the one and two-time correlations are of order one, i.e., 
\begin{equation}
    \kappa_1^E \simeq \kappa_2^E\simeq \mathcal O(1) \ .
    \label{eq:scaling_local}
\end{equation} As a consequence, the energy fluctuations in Eq.~\eqref{k2eth} are always subleading both in the canonical and the microcanonical ensemble:
\begin{subequations}
\label{k2intensive}
\begin{align}
	\frac{\kappa_2^\beta - \aveE{k_2}_{E_\beta}}{\aveE{k_2}_{E_\beta}}  & =  \mathcal O(N^{-1})
\\
	\frac{\kappa_2^{E} - \aveE{k_2}_{E}}{\aveE{k_2}_{E}}  & =  \mathcal O(\Delta^2 N^{-2})\ .
\end{align}    
\end{subequations}
That is, for local observables, the connected equilibrium two-time correlation functions $\kappa_2^E$ do coincide with the respective ensemble average $\overline{k_2^{  }}_{E}$ of the single-eigenstate fluctuations $k_2^{   i}$ for both ensembles with corrections vanishing in the thermodynamics limit as $\sim 1/N$.\\

In contrast, in the case of \emph{collective or extensive} observables (such as sums of local ones $A_{\rm coll}=\sum_r a_r$), the connected fluctuations are usually subleading with respect to the average, namely
\begin{equation}
\label{eq:scaling_extensive}
	\frac{\kappa_2^E}{(\kappa_1^E)^2}= \frac{\langle \hat A^2 \rangle_E 
    - \langle \hat A \rangle_E ^2 }{\langle \hat A \rangle_E ^2 } \simeq \mathcal O(N^{-1}). 
\end{equation}
This implies that in the case of the canonical ensemble, for which $\Delta^{(2)}_{E_\beta}=\mathcal O(N^{-1})$, see Eq.~\eqref{canonicalFluc} the two contributions in Eq.~\eqref{k2eth} are of the same order, i.e.
\begin{subequations}
\label{k2_extensive}
\begin{align}
\frac{\kappa_2^\beta - \aveE{k_2^{}}_{E_\beta}}{\aveE{k_2^{}}_{E_\beta}}  & =  \mathcal O(1). \end{align}
The ensemble-averaged single-eigenstate result does not reproduce the connected equilibrium two-time correlation function even in the thermodynamical limit. 

On the other hand, in the microcanonical case, the corrections arising from energy fluctuations can always be neglected also for extensive observables since the error scales as
\begin{align}	
	\frac{\kappa_2^{E} - \aveE{k_2^{}}_{E}}{\aveE{k_2^{}}_{E}}  & =  \mathcal O(\Delta^2 N^{-1}) \ ,
\end{align}
which can be made arbitrarily small. 
\end{subequations}
Summarizing, \emph{while for local observables, one can always evaluate the equilibrium connected correlations using the ETH result, in the case of extensive observables, this holds only for the microcanonical ensemble.} The difference in the observable scaling of fluctuations will be shown numerically in Sec.~\ref{sec_numer_ETH2} below.  \\

The considerations of this section illustrate the failure of standard ETH averages to reproduce thermal equilibrium two-time correlation functions, i.e., defined with respect to the canonical ensemble, at least in case of extensive observables.
While this is outlined here for two-time correlation functions and standard ETH only, this is also the case for multi-point correlation functions in the context of full ETH. 
A general discussion of equilibrium multi-time correlation functions therefore has to rely on the microcanonical ensemble and will be presented in the remainder of this paper.

\section{Full ETH and Microcanonical Free Cumulants}
\label{sec:fullETH}

Having established that a systematic discussion of multi-time equilibrium correlation functions has to rely on the microcanonical ensemble, we now review the full eigenstate thermalization hypothesis \cite{foini2019eigenstate}, and discuss its connection with free probability \cite{speicher1997free} in the microcanonical setting. 

\subsection{Full Eigenstate Thermalization Hypothesis}

To accurately capture the dynamics of general multi-time correlation functions a general version of the Eigenstate-Thermalization-Hypothesis has been introduced in Ref.\cite{foini2019eigenstate}.
It is an ansatz on the statistical properties of the product of $q$ matrix elements $A_{ij}$ and consequently includes correlations between matrix elements not present in standard ETH. Specifically, the average of products with distinct indices $i_1\neq i_2 \dots \neq i_q$ reads
\begin{subequations}
    \label{GEN_ETH}
\begin{equation}
    \label{ETHq}
    \overline{A_{i_1i_2}A_{i_2i_3}\dots A_{i_{q}i_1}} = e^{-(q-1)S(E^+)} F_{e^+}^{(q)}(\omega_{i_1i_2}, \dots, \omega_{i_{q-1}i_q})
\end{equation}
while, with repeated indices, it shall factorize in the large $N$ limit as
\begin{align}
\label{ETH_conta}
\hspace{-.2cm}
\begin{split}
 \overline{A_{i_1i_2}\dots A_{i_{k-1}i_1}A_{i_1i_{k+1}}\dots A_{i_{q}i_1}}
\\
 =  \overline{A_{i_1i_2}\dots A_{i_{k-1}i_1}} \; 
\overline{A_{i_1i_{k+1}}\dots A_{i_{q}i_1}}    \ .    
\end{split}
\end{align}
\end{subequations}

The ansatz, Eq.~\eqref{ETHq}, is a direct generalization of the standard ETH ansatz, Eq.~\eqref{eq:ETH2}. 
Here, $Ne^+ = E^+=(E_{i_1}+\dots +E_{i_q})/q$ is the average energy, $\vec 
\omega = (\omega_{i_1i_2}, \dots, \omega_{i_{q-1}i_q})$ with $\omega_{ij}=E_i-E_j$ are $q-1$ energy differences and $F_{e^+}^{(q)}(\vec \omega)$ is a smooth function of the energy density $e^+=E^+/N$ and $\vec \omega$. Thanks to the explicit entropic factor, the functions $F^{(q)}_e(\vec \omega)$ are of order one, and they contain all the physical information.\\

\subsection{Free cumulants}

\begin{figure*}[t]
	\includegraphics[width=1\linewidth]{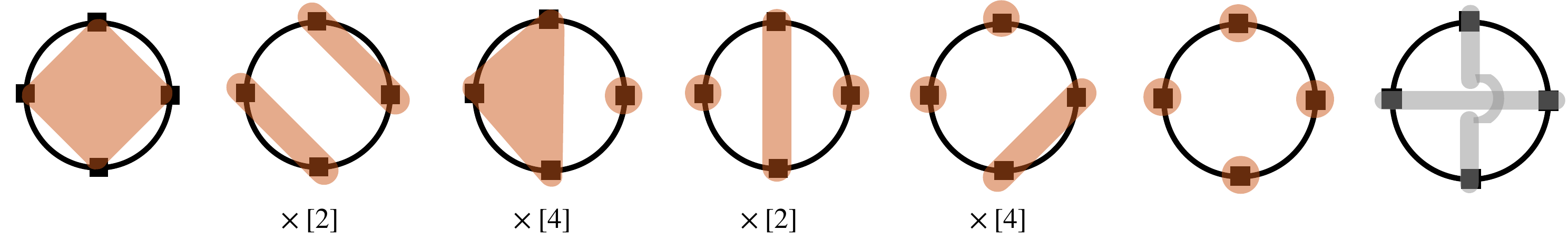}
    \label{fig_P4}
    \caption{Set of all partitions for $n=4$. With the color orange, we represent the non-crossing partitions, while the crossing one is in grey. With $\times [m]$, we denote the $m$ cyclic permutations of that partition, which determines the coefficients appearing in the moment/cumulant formulas in Eq.~\eqref{eq_freecumbeta}.}
 \end{figure*}
 
The full ETH simplifies a lot using the notion of free cumulants \cite{speicher1997free}, which play a central role in free probability and the study of non-commuting random variables.
They are connected multi-point correlation functions, extending the one and two-point functions defined in Eqs.~\eqref{k1k2def} to a higher order. 
Free cumulants are defined recursively and depend on the choice of an expectation value such as Eq.~\eqref{eq:ensemble_average}.
Based on the previous discussion, the obvious choice is that of the expectation value with respect to the microcanonical enemble. 
We hence \emph{define} \emph{microcanonical free cumulants} $\kappa^E_q$ of order $q$ recursively  from the moment-free cumulant formula as
\begin{equation}
    \label{eq_freecumbeta}
    \langle \hat A(t_1) \hat A(t_2) \dots \hat A(t_q) \rangle_{E} = \sum_{\pi\in NC(q)} \kappa^E_\pi \left ( A(t_1),  \dots,  A(t_q) \right )  \ .
\end{equation}

Here, $NC(q)$ denotes the set of non-crossing partitions of a set of $q$ elements. The elements (blocks) of $\pi \in NC(q)$ are disjoint subsets of the $q$-element set, whose union is the whole set and which are subject to a non-crossing condition, see Fig.~\ref{fig_P4} for an example for $NC(4)$.

To each partition $\pi \in NC(q)$ corresponds a microcanonical free cumulant $\kappa^E_\pi$, which itself is a product of microcanoical  free cumulants $\kappa_{|b|}^E$ for each block of $b \in \pi$, with $|b|$ the size of the block.
More precisely, writing the blocks $b$ as $b=(b_1,b_2,\ldots,b_{|b|})$ with $b_1 < b_2 < \ldots < b_{|b|}$ one has
\begin{equation}
     \kappa^E_\pi \left ( A(t_1),  \dots  ,A(t_q) \right ) = \prod_{b\in \pi} \kappa^E_{|b|} \left (A(t_{b_1}), \dots, A(t_{b_{|b|}}) \right ) \ .
\end{equation}

By inverting the implicit definition in Eq.~\eqref{eq_freecumbeta}, the first few free cumulants read
\begin{subequations}
    \label{eq:first_free_cumulants}
    \begin{align}
        \kappa_1^E &  \equiv \langle \hat A\rangle_E\\
        \kappa_2^E(t) & \equiv \langle\hat A(t) \hat A\rangle_E - \langle \hat A\rangle_E^2 \\
         \kappa_3^E(t_1, t_2) & \equiv \langle \hat A(t_1)\hat  A(t_2)\hat  A \rangle_E 
           - \langle\hat  A(t_2) \hat A\rangle_E\langle \hat A\rangle_E  
           \\ & \quad 
          -  \langle \hat A(t_1)\hat A(t_2)\rangle_E \langle \hat A\rangle_E
       - \langle \hat A(t_1)\hat A\rangle_E\langle \hat A\rangle_E 
       \nonumber\\ & \quad           
         + 2 \langle\hat  A \rangle_E^3  \nonumber\\ 
      \kappa_4^E(t_1,t_2,t_3) &  
      \equiv \langle \hat A(t_1) \hat A(t_2) \hat A(t_3) \hat A \rangle_E \\ & \quad \nonumber
      - \braket{ \hat A(t_1) \hat A(t_2)}_E \braket{\hat A(t_3) \hat A}_E 
      \\ & \quad \nonumber
      - \braket{ \hat A(t_1) \hat A}_E \braket{\hat A(t_2) \hat A(t_3)}_E  
      \\
      & - \langle \hat A\rangle_E \Big[
      \langle \hat A(t_2) \hat A(t_3)\hat A \rangle_E
     + \langle \hat A(t_1) \hat A(t_3)\hat A \rangle_E
     \nonumber \\ &\quad 
     + \langle \hat A(t_1)\hat A(t_2) \hat A \rangle_E
     +   \langle \hat A(t_1) \hat A(t_2)\hat A(t_3) \rangle_E
  \nonumber  \\
    & \quad +
\langle \hat A\rangle_E\Big (
   2\langle \hat{A}(t_1) \hat{A} \rangle_E 
 + 2 \langle \hat{A}(t_3) \hat{A} \rangle_E 
\nonumber \\ & \quad 
\quad + 2 \langle \hat{A}(t_2) \hat{A}(t_3) \rangle_E 
+ 2 \langle \hat{A}(t_1) \hat{A}(t_2) \rangle_E 
\nonumber \\  &  \quad 
 \quad  + \langle \hat{A}(t_2) \hat{A} \rangle_E + \langle \hat{A}(t_1) \hat{A}(t_3) \rangle_E \Big )
   - 5 \langle \hat A\rangle^3
      \Big ]\nonumber
      \\ &
      \ldots \nonumber  \ .
    \end{align}
\end{subequations}
They differ from classical cumulants for order $q\geq 4$.
We simplified the notation by writing
$\kappa^E_\pi \left ( t_1, t_2, \dots t_q \right ) = \kappa^E_\pi \left ( A(t_1) A(t_2) \dots  A(t_q) \right )$
and by using \emph{time translational symmetry}, i.e. $\langle\hat  A(t_1)  \dots \hat A(t_q) \rangle_E=\langle \hat A(t_1-t_q)  \dots \hat A(0) \rangle_E$ and  $\kappa^E_\pi \left ( t_1,  \dots t_q \right ) = \kappa^E_\pi \left ( t_1-t_q,\dots 0 \right )$. For $q=1$ this reads ${\kappa_1^{E}(t)=\kappa_1^E(0)= \braket{\hat A}_E}$.
In particular, the above corresponds to Eqs.~\eqref{k1k2def} for $q=1$ and $q=2$.

We recall that this is just an implicit definition of cumulants in terms of moments, which can be defined in principle also for integrable or non-ergodic systems and for any choice of expectation value. 
We will now discuss how this definition simplifies the discussion of the general ETH, which, in turn, implies a particularly simple form for the microcanonical free cumulants.

\subsection{... and ETH}

The decomposition in free cumulants allows for a systematic characterization of the multi-time correlation function and their interplay with the full ETH ansatz~\eqref{ETHq}.
The full ETH ansatz implies a particularly simple, non-recursive form for the free cumulant $\kappa^{E}_{q}$ defined recursively in Eq.~\eqref{eq_freecumbeta}.
This leads to the main result of the Free Probability approach to ETH:
as we will now show, the microcanonical free cumulants of ETH-obeying systems are given by the ETH averages \cite{pappalardi2022eigenstate}
\begin{equation}
    \label{ETH_kq}
 \kappa^{E}_{q}(\vec t) = \aveE{k^{{  }}_{q}(\vec t)}_E  +\mathcal O(\Delta^2, N^{-1}) \ .
\end{equation}
For $q=1$ and $q=2$ this was demonstrated in Sec.~\ref{sec_fluctua}. 
For larger $q$ we will establish the above by showing that the ETH average $ \aveE{k^{{  }}_{q}(\vec t)}_E$ obeys the same recursion relation as $\kappa_q^E$.
To be precise, the right hand side of Eq.~\eqref{ETH_kq} is determined by $k^{i}_{q}(\vec t)$, which generalizes the single eigenstate definition of Eqs.~\eqref{k1k2_singleE} and is given by the sum over ``simple loops'' (non-repeated indices arranged on a loop)
\begin{align}
    \label{kq_nnonsmooth}
   & k^{{i_1}}_{q} \left (  t_1,  \dots, t_q \right )  \nonumber \\
   &= \sum_{i_2 \neq \dots \neq i_q: \neq i_1} A_{i_1i_2}\dots A_{i_{q}i_1}e^{i t_1 \omega_{i_1i_2} +\dots i t_q \omega_{i_{q}i_1}}\,.
\end{align}
Here, the summation runs over distinct indices $i_n$ with $n\neq 1$, which are additionally distinct from $i_1$.
The smoothed averaged result in Eq.~\eqref{ETH_kq} is then computed as
\begin{align}
    \label{freekETH}
  & \aveE{ k^{{}}_{q} \left (  t_1, \dots, t_q \right )}_E \\ & = 
   \sum_{i_1\neq i_2 \neq \dots \neq i_q} \frac{p_{E}(E_{i_1})}{Z_E} A_{i_1i_2}\dots A_{i_{q}i_1}e^{i t_1 \omega_{i_1i_2} +\dots i t_q \omega_{i_{q}i_1}} \ .\nonumber 
\end{align}
In particular the ETH averages relate to the Fourier transform, $\text {FT} [\bullet]= \int d\vec \omega e^{-i \vec \omega \cdot \vec t} \bullet$, of the smooth ETH functions $F^{(q)}_{e}(\vec \omega)$ in Eq.~\eqref{ETHq} via
\begin{align}
    \label{freekETH_Fourier}
   \aveE{ k^{{}}_{q} & \left (  t_1, t_2, \dots, t_q \right )}_E 
   \\ &  = \text {FT} \left [ F^{(q)}_{e}(\vec \omega) e^{-S' \vec \omega \cdot \vec \ell_q} \delta(\omega_1 +\omega_2 + \dots +\omega_q)\right ] \ . \nonumber
\end{align}
The exponent with 
$\vec \ell_q = \left (\frac{q-1}{q}, \dots, \frac 1q , 0\right )
$ corresponds to a generalization of the fluctuation-dissipation theorem in the microcanonical ensemble \cite{haehl2017thermal}, See the derivation in Sec.~\eqref{app_derive}.

\emph{This result shows that all the correlations of the general ETH \eqref{ETHq} are encoded precisely in the microcanonical free cumulants. } Furthermore, it gives an immediate tool to compute higher-order correlation functions directly in terms of the ETH correlations, by substituting in Eq.~\eqref{eq_freecumbeta} directly the ETH result $ \aveE{k^{{}}_{q}(\vec t)}_E$.
For instance, in the example of four-point functions, the validity of the ETH ansatz implies
\begin{widetext}
\begin{align}
    \begin{split}
        \langle \hat A(t_1) \hat A(t_2) \hat A(t_3) \hat A(t_4) \rangle_E = & 
         \aveE{k_4^{}(t_1, t_2, t_3, t_4)}_E  
         +  \aveE{k_2^{}(t_1,t_2)}_E  \aveE{k^{}_2(t_3, t_4)}_E
        +  \aveE{k_2^{}(t_1,t_4)}_E  \aveE{k^{}_2(t_2, t_3)}_E
        \\ & 
        +  \aveE{k^{{}}_1}_E \Big [
         \aveE{k^{}_3(t_1, t_2, t_3)}_E  
        +  \aveE{k^{{}}_3(t_1, t_3, t_4)}_E
        +  \aveE{k^{{}}_3(t_1, t_2, t_4)}_E 
        +  \aveE{k^{{}}_3(t_2, t_3, t_4)}_E \\
        & \quad\quad \quad\quad \quad\quad   +  \aveE{k^{}_1}_E  \aveE{k^{}_2(t_1, t_3)}_E
        +  \aveE{k^{{}}_1}_E  \aveE{k^{{}}_2(t_2, t_4)}_E + \aveE{k^{{}}_1}_E
        ^3 
        \\
         & \quad\quad \quad\quad \quad\quad  
       +   \aveE{k^{{}}_1}_E  \aveE{k^{{}}_2(t_1, t_2)}_E  
        +  \aveE{k^{{}}_1}_E  \aveE{k^{{}}_2(t_1, t_4)}_E 
         \\
         & \quad\quad \quad\quad \quad\quad  
        +  \aveE{k^{{}}_1}_E  \aveE{k^{{}}_2(t_2, t_3)}_E 
        +  \aveE{k^{{}}_1}_E  \aveE{k^{{}}_2(t_3, t_4)}_E
        \Big ]\ ,
    \end{split}
    \label{eq:4point_correlator}
\end{align}
\end{widetext}
which follows from inserting Eq.~\eqref{ETH_kq} into the general result Eq.~\eqref{eq_freecumbeta}.

\subsubsection{Ground of validity}
\label{sec_validity}
The validity of Eq.~\eqref{ETH_kq} was shown for local observables in the canonical ensemble in Ref.~\cite{pappalardi2022eigenstate} with a correction $\mathcal O(N^{-1})$ instead of $\mathcal O(\Delta^2)$. In the following we now discuss the case of the microcanonical ensemble, focusing in particular on extensive observables. \\

\begin{figure*}[t]
\includegraphics[width=1\linewidth]{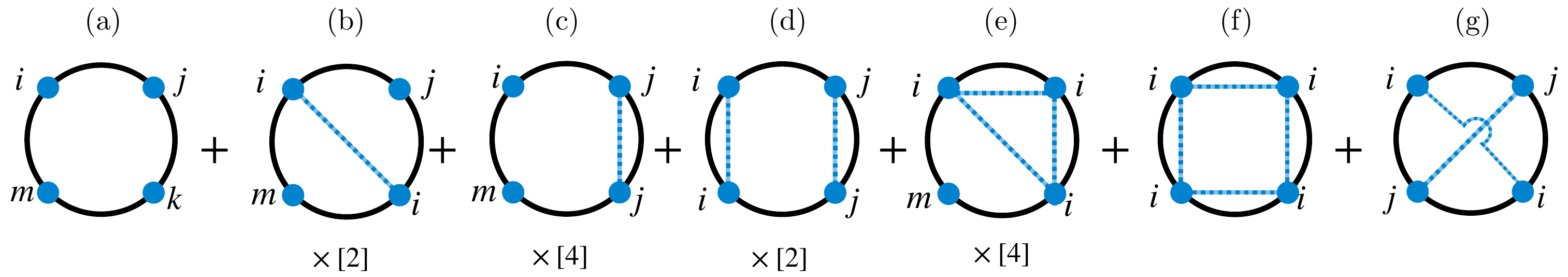}
    \caption{ETH indices diagrams: bookeeping of ETH matrix elements for $n=4$. Matrix elements $A_{ij}$ lie on the outer black lines connecting two vertices, which represent the corresponding energy indices. Blue dots represent different indices, and the edges connecting two or more dots represent a contraction among them. }
    \label{fig_ethDiag}
 \end{figure*}
 
To show the decomposition of the four-time correlator~\eqref{eq:4point_correlator} or more general for arbitrary multi-time correlation function we write them as
\begin{align}
\label{eq_Aq}
   & \langle \hat A(t_1)  \hat A(t_2) \dots  \hat A(t_{q-1})
   \hat A \rangle_E 
    \\ & = \sum_{i_1i_2\dots i_q}\frac{p_E(E_{i_1})}{Z_E}{A_{i_1i_2}A_{i_2i_3}\dots A_{i_{q}i_1}} e^{i t_1 \omega_{i_1i_2} +\dots i  t_{q-1} \omega_{i_{q-1}i_q}}\ ,\nonumber
\end{align}
and systematically study all the contributions of the different matrix elements from all the possible index contractions.
One can do this \emph{diagrammatically} by introducing \emph{ETH indices diagrams}. As an example, we consider four-point functions that we illustrate pictorially in Fig.~\ref{fig_ethDiag}.  Products $A(t_1)_{ij}A(t_2)_{jk}A(t_3)_{km}A_{mi}$, with $A(t)_{ij}=A_{ij}e^{i t \omega_{ij}}$, are represented on a loop with four vertices $i,j,k,m$, depicting energy eigenstates.  The contractions between two or more indices are represented by lines that connect the vertices. The blue dots indicate that the indices are all different. For instance, suppressing the time dependence in the notation, the first diagram represents $A_{ij}A_{jk}A_{km}A_{mi}$, the second $A_{ij}A_{ji}A_{im}A_{mi}$, the third $A_{ij}A_{jj}A_{jm}A_{mi}$  with all distinct indices, and so on. 
With $\times [n]$ we indicate that there are $n$ cyclic permutations of the indices.

One recognizes that there are two types of index diagrams: 
\begin{itemize}
    \item \emph{Crossing ones} in which the lines cross, i.e.~(g) in Fig.~\ref{fig_ethDiag}.
    \item \emph{Non-crossing} ones -- in which the polygons created by the groups of identical indices do not cross. This includes (a) the simple loops with all different indices as well as the diagrams (b-f) in Fig.~\ref{fig_ethDiag};
\end{itemize}

Let us denote by $\kappa_{(x)}$ the sum of the matrix elements associated to the ETH diagram $(x)$, in reference to Fig.~\ref{fig_ethDiag}, for instance $\kappa_{(a)}(t_1, t_2, t_3) =  \sum_{i\neq j\neq k \neq m} \frac{p_{E}(E_i)}{Z_E}  
        A(t_1)_{ij}A(t_2)_{jk}  A(t_3)_{km} A_{mi}$. To compute the correlation function $\langle A(t_1) A(t_2) \dots A(t_q) \rangle_E $ in Eq.~\eqref{eq_Aq}, one shall sum over all such ETH diagrams with their respective multiplicity. 
        This calculation, which may seem challenging for generic $q$, simplifies using the language of Free Probability. 
In fact, the general ETH ansatz in Eqs.\eqref{GEN_ETH} in the large $N$ limit implies that
\begin{enumerate}
    \item \emph{Crossing diagrams are suppressed with the inverse of the density of states.} 
        \item All \emph{non-crossing diagrams} yield a finite contribution with \emph{factorization of non-crossing diagrams into products of irreducible simple loops}. 
\end{enumerate}
The first property implies that crossing diagrams can be neglected when calculating higher-order correlation functions. Thus, the combinatorics of multi-time correlation functions are the same as the ones known in Free Probability: all the contributions to multi-time correlations have to be found in non-crossing partitions, which themselves factorize into the ETH averages.

The second property implies the factorization of the non-crossing diagrams, for instance, that the diagrams (b-f) in Fig.~\ref{fig_ethDiag} can be cut along the blue line. This implies that the non-crossing ETH diagrams can be read as the ``dual''  of non-crossing partitions $\pi$ in which every element of the set is not associated with an observable [for instance (b-f) in Fig.~\ref{fig_ethDiag} 
 can be read as (a-f) in Fig.~\ref{fig_P4}]. 
This shows that the microcanonical free cumulants are given at every order by the smoothed average of the simple loops, see Eq.~\eqref{ETH_kq}. \\

\subsubsection{Derivation}
Let us now outline how the properties 1. and 2. arise.

\emph{Property 1.} The suppression of crossing diagrams originates from entropic counting: crossing diagrams do not have enough free indices to counterbalance the matrix elements' entropic dependencies. 
For instance, in the $q=4$ example in reference to Fig.~\ref{fig_ethDiag}(g), reads
  \begin{equation}
  \label{full_ETH2}
   \kappa_{(g)}(t) =  \sum_{i\neq j} \frac{p_{E}(E_i)}{Z_E}  e^{\ii\omega_{ij} t} {|A_{ij}|^4} 
 \simeq \frac{1}{Z_E} \simeq e^{- S(E)} \ ,
\end{equation}
where we substituted the ETH scaling of the matrix elements $|A_{ij}|^4\sim e^{-2S}$, while $\sum_{i\neq j}\sim e^{2S}$. \\

\emph{Property 2.} The factorization of the non-crossing diagrams follows from the ETH smoothness of the individual matrix elements, which are averaged on a peaked equilibrium ensemble. In other words, non-crossing diagrams factorize whenever we can substitute the smoothed average with their arguments at the averaged energy, i.e. Eq.~\eqref{micro_ave}. This argument is exactly the same as we described in detail in Section \ref{sec:scaling} for $ \aveE{A_{ii}^2}_E =  \aveE{A_{ii}}_E^2 +\mathcal O(\Delta^2/N^2)$, that we now generalize to many indices observables. This is the property that fails for extensive observables in the case of the canonical ensemble.

For instance, in the $q=4$ example, we list the contributions from the ETH diagrams in Fig.~~\ref{fig_ethDiag}. 
To this end, we note that the $[n]$ cyclic permutations of indices yield the same contribution to the correlation functions only in the large $N$ limit.
We distinguish two sub-classes of non-crossing diagrams: the first one is given by averages of smooth functions all at the same energy $e_i$. Their factorization is valid at an order $\mathcal O(\Delta^2)$. Specifically:

\begin{widetext}
\begin{subequations}
    \label{eq:facto_eth_diagrams}
    \begin{align}
        \kappa_{(a)}(t_1, t_2, t_3) & =  
        \sum_{i\neq j\neq k \neq m} \frac{p_{E}(E_i)}{Z_E}  
        A(t_1)_{ij}A(t_2)_{jk}  A(t_3)_{km} A_{mi}
        =  \aveE{k_4(t_1, t_2, t_3)}_E
        \\
        \label{cactusb}
    \kappa_{(b)}(t_1, t_2, t_3) & = \sum_{i\neq j \neq m} \frac{p_{E}(E_i)}{Z_E}  
        A(t_1)_{ij}A(t_2)_{ji}  A(t_3)_{im} A_{mi}
    =  \aveE{k_2(t_1, t_2)}_E\;  \aveE{k_2(t_3, 0)}_E + \mathcal O(\Delta^2)
        \\ 
         \kappa_{(c)}(t_1, t_2, t_3) &  = \sum_{i\neq j \neq k} \frac{p_{E}(E_i)}{Z_E}  
        A(t_1)_{ii}A(t_2)_{ik}  A(t_3)_{km} A_{mi}
            =  \aveE{k_1}_E\;  \aveE{k_3(t_2, t_3, 0)}_E + \mathcal O(\Delta^2)
        \\ 
    \kappa_{(e)}(t_1, t_2, t_3) &  = \sum_{i\neq j} \frac{p_{E}(E_i)}{Z_E}   
    A(t_1)_{ii}A(t_2)_{ii}  A(t_3)_{ij} A_{ji}  =   \aveE{k_1}_E^2\;  \aveE{k_2(t_3, 0)}_E 
        + \mathcal O(\Delta^2)
        \\ 
             \label{cactusf}
    \kappa_{(f)}(t_1, t_2, t_3) & = \sum_{i} \frac{p_{E}(E_i)}{Z_E}  
        A(t_1)_{ii}A(t_2)_{ii}  A(t_3)_{ii} A_{ii}
    =  \aveE{k_1}_E^4+ \mathcal O(\Delta^2)
    \end{align}
\end{subequations} 
\end{widetext}

Such factorization holds in the large $N$ limit and is derived using the ETH ansatz \eqref{ETH_conta} combined with the microcanonical average in Eq.~\eqref{micro_ave}. The above class of diagrams, given by smoothed averages of products of simple loops all at the same energy, yields for instance
\begin{align}
\begin{split}
\label{kb}
    \kappa_{(b)}&(t_1, t_2, t_3) 
    \\ & = \sum_{i} \frac{p_{E}(E_i)}{Z_E}  \sum_{j: j\neq i}
        A(t_1)_{ij}A(t_2)_{ji}   \sum_{m: m\neq i} A(t_3)_{im} A_{mi} 
        \\ & =  \aveE{k_2^{{ }i}(t_1,t_2) k_2^{{ }i}(t_3, o)}_E 
 \\ & 
    =  \aveE{k_2(t_1, t_2)}_E\;  \aveE{k_2(t_3, 0)}_E + \mathcal O(\Delta^2)
    \end{split}\ ,
\end{align}
where one first identifies the smooth average at the same energy density $E_i/N$ over the simple loops and then uses the microcanonical average in Eq.~\eqref{micro_ave}.
These are microcanonical smoothed averages of products of simple loops $\kappa_n^{  i}$ all the same energy, which are smooth in energy density $e_i$ due to ETH. Hence, they factorize in the products of averages at the energy density $e$, generalizing Eq.~\eqref{facto2}.\\

The second sub-class of diagrams is given by smoothed averages of simple loops at ``slightly'' different energies compared to $e_i$. In this case, we use the fact that the smooth ETH functions defined in Eqs.~\eqref{eq:ETH2} are functions of the energy densities $e_k$; as such, they can always be tailored expanded around $e_i$ with an error $\mathcal O(N^{-1})$, i.e. $e_k=e_i -\omega_{ik}/N$. 
This class also includes possible inequivalent cyclic permutations of the index diagrams in Fig.~\ref{fig_ethDiag}.
In the example $q=4$, there are two inequivalent permutations for the diagrams (b), (c) and (e), whose contributions differ by a factor of the order $1/N$.
We refer to these diagrams as, e.g., $\kappa_{(x)}$ and $\kappa_{(x^\prime)}$, respectively. 
For the other diagrams, all permutations lead to identical contributions to the correlation functions at every $N$. This other sub-class of diagrams factorizes as

\begin{widetext}
\begin{subequations}  \label{eq:facto_eth_diagrams2}
    \begin{align}
         \label{cactusb1}
    \kappa_{(b')}(t_1, t_2, t_3) & = \sum_{i\neq j \neq m} \frac{p_{E}(E_i)}{Z_E}  
        A(t_1)_{ij}A(t_2)_{jk}  A(t_3)_{kj} A_{ji}
    =  \aveE{k_2(t_1, 0)}_E\;  \aveE{k_2(t_2, t_3)}_E + \mathcal O(\Delta^2, N^{-1})
         \\
          \kappa_{(c^\prime)}(t_1, t_2, t_3) &  = \sum_{i\neq k \neq m} \frac{p_{E}(E_i)}{Z_E}  
        A(t_1)_{ij}A(t_2)_{jj}  A(t_3)_{jk} A_{ki}
            =  \aveE{k_1}_E\;  \aveE{k_3(t_1, t_3, 0)}_E + \mathcal O(\Delta^2, N^{-1})
         \\
    \kappa_{(d)}(t_1, t_2, t_3) &  = \sum_{i\neq j} \frac{p_{E}(E_i)}{Z_E}  
        A(t_1)_{ij}A(t_2)_{jj}  A(t_3)_{ji} A_{ii}  =   \aveE{k_1}_E^2\;  \aveE{k_2(t_1, t_3)}_E + \mathcal O(\Delta^2, N^{-1})
                  \\ 
    \kappa_{(e^\prime)}(t_1, t_2, t_3) &  = \sum_{i\neq j} \frac{p_{E}(E_i)}{Z_E}  
     A(t_1)_{ij}A(t_2)_{jj}  A(t_3)_{jj} A_{ji}  =   \aveE{k_1}_E^2\;  \aveE{k_2(t_1, 0)}_E
       + \mathcal O(\Delta^2, N^{-1})
    \end{align}
\end{subequations}
\end{widetext}
To show the above factorization, we illustrate the example of Eq.~\eqref{cactusb1}, which reads
\begin{align}
\label{kb1exp}
\begin{split}
    \kappa_{(b')}(t_1, t_2, t_3) & = \sum_{i} \frac{p_{E}(E_i)}{Z_E}  
        \sum_{j:\neq i}A(t_1)_{ij} A_{ji} \sum_{k: k\neq j}A(t_2)_{jk}  A(t_3)_{kj} \\
        & =  \sum_{i} \frac{p_{E}(E_i)}{Z_E}  
        \sum_{j:\neq i}A(t_1)_{ij} A_{ji}\,\, k^{   j}_2(t_2, t_3) 
\end{split}
\end{align}
we can now use that within ETH $ k^{j}_2(t_2, t_3)$ is a smooth function of the energy \emph{density} $e_j = E_j/N$, hence we can expand $e_j = e_i + \omega_{ji}/N$ around energy $e_i$ as
\begin{equation}
    k^{   j}_2 (t_2, t_3) 
    =  k^{i}_2(t_2, t_3) 
    + \mathcal O(\omega_{ji} N^{-1})\ .
\end{equation}
The contribution $\mathcal O(\omega_{ji} N^{-1})$ never becomes relevant, because it is multiplied by $A(t_1)_{ij} A_{ji}\sim F^{(2)}_{e^+}(\omega_{ij})$ that is assumed to decay rapidly to zero at large frequencies \cite{dalessio2016from}. Plugging this back into Eq.~\eqref{kb1exp} one has
\begin{align}
\label{kb1exp_}
\begin{split}
    \kappa_{(b')}(t_1, t_2, t_3) 
        & =  \sum_{i} \frac{p_{E}(E_i)}{Z_E}  
        k^{i}_2(t_1, 0)\,\, (k^{i}_2(t_2, t_3) 
    + \mathcal O(N^{-1}))
    \\ &  =  \aveE{k_2(t_1, 0)\, k_2(t_2, t_3)}_E  + \mathcal O(N^{-1})
   \\ &  =
     \aveE{k_2(t_1, 0)}_E\;  \aveE{k_2(t_2, t_3)}_E + \mathcal O(\Delta^2, N^{-1})
\end{split}\ .
\end{align}
Consequently, the diagram $\kappa_{(b^\prime)}$ factorizes in the same way as the $\kappa_{(b)}$ and in Eq.~\eqref{kb} up to a correction of order $1/N$, i.e.
$ \kappa_{(b^\prime)}(t_1, t_2, t_3)=\kappa_{(b)}(t_1, 0, t_2-t_3) + \mathcal O(N^{-1})\ .$
A similar derivation yields analogous statements also for the diagrams (c) and (e).
These arguments apply also the factorization of diagram (d), which differs from diagram (e) up to $1/N$ corrections
$ \kappa_{(d)}(t_1, t_2, t_3)=\kappa_{(e)}(0, 0, t_3-t_1) + \mathcal O(N^{-1})\ ,$ see derivation in App.~\ref{app:factorization}. This also leads to an order $O(\Delta^2, 1/N)$ correction to the factorized result. 

We note that the existence of these two distinct classes of diagrams is a consequence of choosing a distribution $p_E(E_i)$ peaked around some energy, e.g., the microcanonical or the canonical distribution. In the case of the infinite temperature state, both classes of diagrams yield, in fact, the same results.

Eventually, the above reasoning implies, that the ETH averages coincide with the cumulants as obtained from Free Probability. In contrast to the latter, they are defined by the explicit formula~\eqref{freekETH} and hence provide a convenient tool to systematically study multi-time correlation functions.

\section{Numerical evaluation}
\label{sec_num}
\begin{figure}[]
	\centering
	\includegraphics[width=8.5cm]{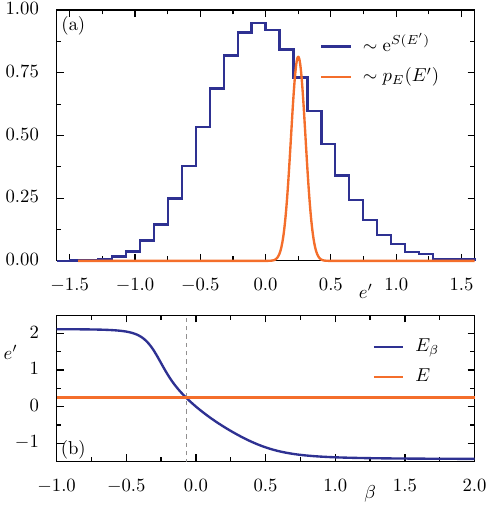}
	\caption{(a) Density of states vs energy density $e^\prime$ for $N=18$ compared with the  microcanonical window Eq.~\eqref{DOS} at $E=e N = 0.25 N$ with width $\Delta=1$.
 (b) Canonical Energy $E_\beta = \langle H \rangle_\beta$ vs. inverse temperature $\beta$.}
	\label{fig_DOS}
\end{figure}

\begin{figure*}
	\centering
    \includegraphics[width=17.cm]{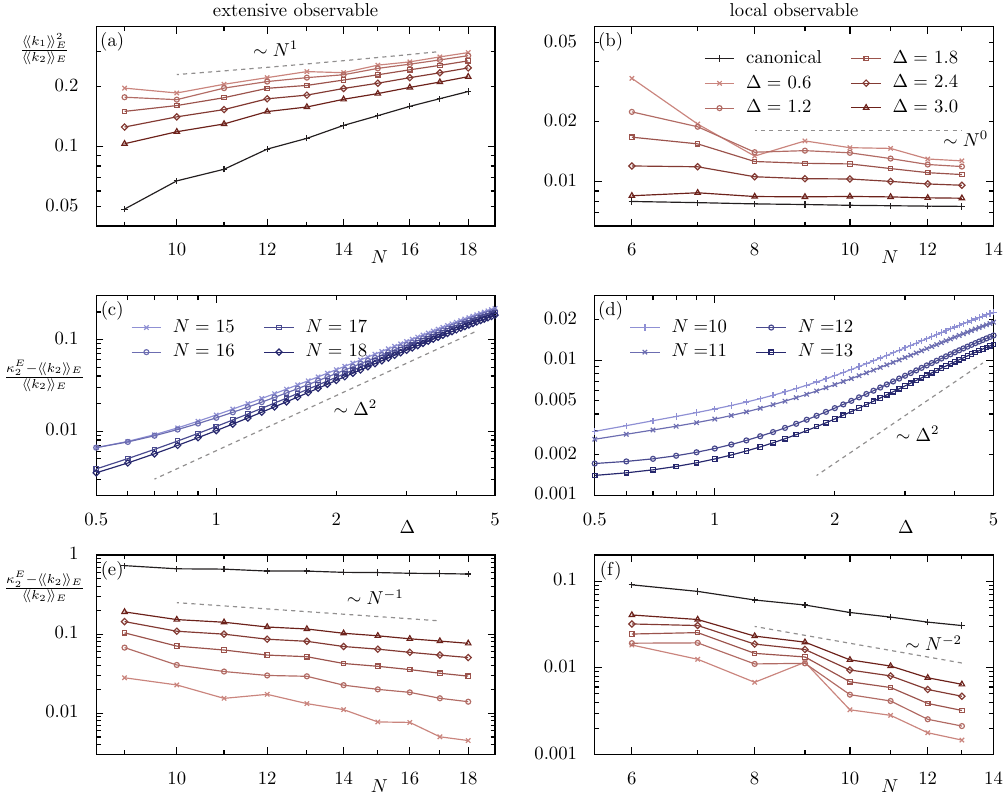}
	\caption{Fluctuations of diagonal matrix elements for extensive (left) and local (right) observable $\hat A$ and $\hat a$, respectively, at microcanonical energy $E=0.25N$. (a,b) Scaling of fluctuations vs. system size $N$ for different microcanonical widths $\Delta$ and for the canonical ensemble. (c-f) Relative deviations between $\kappa_2^E$ and $ \aveE{k_2}_E$ vs. $\Delta$ (c,d)  and vs. $N$ (e,f) for different $\Delta$ and $N$, respectively. Dashed gray lines correspond to the predicted scaling behavior. All plots are in log-log scale. }
	\label{fig:diagonal_fluctuations}
\end{figure*}

In this section we test the above theoretical considerations numerically in a chaotic system.
To this end we consider the non-integrable Ising chain of lenght $L$ subject to transverse and longitudinal fields described by 
\begin{equation}
\label{ising}
\hat H = \sum_{i=1}^N w  \hat \sigma_i^x + \sum_{i=1}^N h \hat \sigma_i^z + \sum_{i=1}^{N}J \hat \sigma_i^z  \hat \sigma_{i+1}^z \ ,
\end{equation}
with $\sigma^\alpha_i$ the Pauli matrix in direction $\alpha = x,y,z$ acting on site $i$. 
We fix the magnetic fields as $w=\sqrt 5/2$, $h=(\sqrt 5+ 5)/8$, ensuring ergodic dynamics and measure time in units of $J$.
We impose periodic boundary conditions, $N+1=1$, leading to translation and space reflection symmetry. 

\subsection{Standard ETH: two-point functions in the different ensembles}
\label{sec_numer_ETH2}

We first aim at investigating the effects of entropic contributions to canonical averages and the ETH cumulants for both local and extensive observables. We contrast them with their microcanonical counterpart and the respective depencence on width $\Delta$ of the microcanonical energy window.
We here want to demonstrate the scaling behavior predicted by Eq.~\eqref{k2intensive} and \eqref{k2_extensive}.
We choose the total magnetization in $x$ direction 
\begin{equation}
    \hat A = \sum_{i=1}^N \hat \sigma^x_i
    \label{eq:ext_observable}
\end{equation}
as an example of an extensive observable and the local observable
\begin{equation}
    \hat  a = \hat \sigma^x_{\lfloor N/2 \rfloor}
    \label{eq:loc_observable}
\end{equation}
in the middle of the chain.
For the extensive observable~\eqref{eq:ext_observable}, which shares the symmetries of the Hamiltonian, we restrict ourselves to the zero-momentum and even parity sector.
In contrast, for the local observable~\eqref{eq:loc_observable} we consider the full Hilbert space.
To compare different system sizes we fix energy density $e=0.25$ leading to the microcanonical energies $E=Ne$.
This is close enough to the center of the spectrum to ensure that the microcanonical energy shell contains enough states to give meaningful statistical averages. This is illustrated in Fig.~\ref{fig_DOS}(a).
On the other hand, it is sufficiently far away from the center of the spectrum, such that the corresponding canonical state is not the infinite temperature state, i.e., inverse temperature $\beta \neq 0$.
In fact we fix inverse temperature $\beta$ by demanding $E=E_\beta=\mathrm{tr}\left(\hat \rho_{\beta} \hat{H} \right)$ with $\hat \rho_{\beta}$ the canonical density matrix, see. Fig.~\ref{fig_DOS}(b).
Note that for our choice of $E$ one has negative temperature, $\beta<0$.

Following the arguments of Sec.~\ref{sec:scaling} we aim at confirming the scaling predicted by Eqs.~\eqref{eq:scaling_extensive} and \eqref{eq:scaling_local} for extensive and local observables, respectively.
We illustrate the respective scaling of $(\kappa^E_1)^2/\kappa_2^E(0) \approx  \aveE{k_1}_E^2 /  \aveE{k_2(0)}_E$ with system size $N$ in Fig.~\ref{fig:diagonal_fluctuations}(a,b) for extensive and local observables with respect to both the microcanonical state with various choices for $\Delta$ and the canonical state.
In all cases we observe the expected scaling behavior, namely small, sub-extensive fluctuations $\sim N^{-1}$ for extensive observables and order one fluctuations $\sim N^{0}$ for local observables. 

Additionally, we investigate the difference of the second free cumulant $\kappa_2^E(t)$ from the ETH result $ \aveE{k_2(t)}_E$
determined by Eq.~\eqref{k2eth} normalized by $ \aveE{k_2(t)}_E$. That is, we numerically evaluate
$(\kappa_2^E- \aveE{k_2(t)}_E)/ \aveE{k_2(t)}_E$ at $t=0$.
Again we consider averages both with respect to the canonical and the microcanonical ensemble.
In the latter the corrections should scale as $\sim \Delta^{(2)}_E\sim \Delta^2$ with the width of the microcanonical energy shell $\Delta$.
This is illustrated in Fig.~\ref{fig:diagonal_fluctuations}(c,d) for both the extensive and the local observable. While the predicted scaling is  clearly seen in case of the extensive observable for all $\Delta$ the local observable shows deviations. Those are particularly pronounced at small $\Delta$, which we attribute to the finite number of states within the microcanonical energy shell.
Even though the observed scaling behavior differs slightly from the prediction, our numerical results still confirm that entropic contributions to microcanonical averages can be suppressed by choosing $\Delta$ sufficiently small, leading to ${\Delta^{(2)}_E}=\Delta^2$.

Ultimately, in Fig.~\ref{fig:diagonal_fluctuations}(e,f), we demonstrate the scaling with system size $N$.
In the microcanonical ensemble, we observe the latter to scale $\sim N^{-1}$ and $N^{-2}$ for the extensive and local observable, respectively.
This clearly indicates that the second ETH cumulant approaches the second free cumulant in the thermodynamic limit.
We contrast this result with the one canonical ensemble one (black in the plots) where the entropic contributions yield an additional factor of $N$, leading to a scaling behavior $\sim N^{0}$ and $N^{-1}$, respectively, as is also visible in Fig.~\ref{fig:diagonal_fluctuations}(e,f).
While this is unproblematic for the local observables, it clearly indicates that for extensive observables, deviations between the two types of cumulants remain of order one, even in the thermodynamic limit.

\subsection{Full ETH: four point functions and free cumulants}

While the case of local observables is less sensitive to the choice of the ensemble and is treated in Ref.~\cite{ourLetter}, the extensive energy fluctuations in the canonical ensemble spoil the intriguing connection between full ETH and free probability for extensive observables.
Thus, we henceforth focus on extensive observables in the microcanonical ensemble, for which we fix the width $\Delta=1$ in the following.
The corresponding probability distribution is illustrated in Fig.~\ref{fig_DOS}(a).

We aim for testing the predictions of full ETH in this setting.
The most striking consequence is the decomposition of correlation functions, Eq.~\eqref{eq:4point_correlator}, which we test for out-of-time-order correlators (OTOC)
\begin{equation}
    \langle \hat A(t) \hat A(0) \hat A(t) \hat A(0)\rangle_E \ .
\end{equation}
Subsequently, we also test the factorization property, Eq.~\eqref{ETH_conta}, the suppression of crossing contributions, Eq.~\eqref{full_ETH2} and the smooth frequency dependence of the ETH functions $F^{(q)}_{e^+}(\vec \omega)$, Eq.~\eqref{ETHq}, or equivalently the ETH cumulants $ \aveE{k_q(\vec \omega)}_E$.

We simplify the discussion of the OTOC by shifting $\hat A$, Eq.~\eqref{eq:ext_observable} by the first cumulant, i.e., we consider $\hat A_0 = \hat A -  \aveE{k_1(A)}_E \hat {\mathbf{1}}$. This results in $ \aveE{k_1(A_0)}_E=0$, while leaving all other ETH cumulants invariant. 
The OTOC and its decomposition is then given by Eq.~\eqref{eq:4point_correlator} with $t_1=t_3=t$ and $t_2=t_4=0$ and all terms proportional to $k_1$ vanishing.
In particular, we only need to check the suppresion of the crossing diagram $\kappa_{(g)}$ as well as factorization of the diagrams $(b)$ and $(b^\prime)$. We focus here on $(b)$ only and investigate the factorization of $(b^\prime)$ as well all other diagrams for the original observable $A$ in the subsequent section. 

\begin{figure}[]
	\centering
	\includegraphics[width=8.5cm]{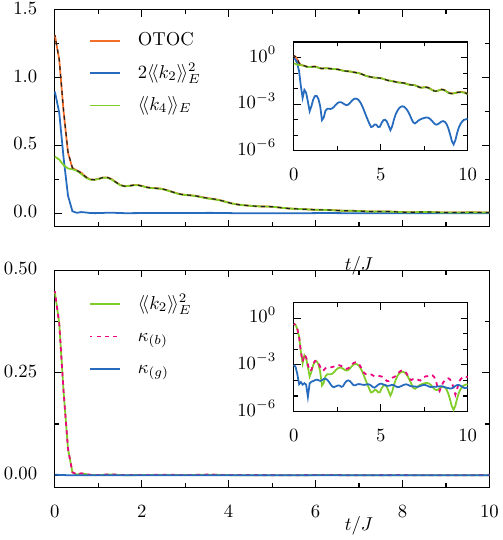}
	\caption{(a) OTOC and ETH cumulants for the observable $\hat{A}_0$ with $\hat A$ defined by Eq.~\eqref{eq:ext_observable} for system size $N=16$ at microcanonical energy $E=0.25N$ and width $\Delta = 1$. The decomposition into free cumulants, Eq.~\eqref{eq:4point_correlator}, is depicted as dashed black line. (b) Factorization of  $\kappa_{(b)}(t) \approx  \aveE{k_2(t)}_E^2$ and crossing contribution $\kappa_{(g)}(t)$.
    The insets show the same data on a semi-logarithmic scale. }
	\label{fig:1}
\end{figure}

The dynamics of the OTOC is depicted in Fig.~\ref{fig:1}(a), which impressively confirms the above decomposition on all time scales shown.
As $ \aveE{k_2(t)}_E$ quickly decays, the OTOC is dominated by the fourth cumulant $ \aveE{k_4(t)}_E$ for all but the shortest times.
This highlights the importance of the correlations captured by $ \aveE{k_4(t)}_E$ in the full ETH ansatz.
The accuracy of the decomposition is facilitated by the factorization of the ETH diagram $\kappa_{(b)}(t)$  according to Eq.~\eqref{cactusb}
as well as the suppression of crossing contribution $\kappa_{(g)}(t)$.
This is illustrated in Fig.~\ref{fig:1}(b), which confirms the factorization for initial times, covering in particular the times, for which $\kappa_{b}(t)$ gives a non-negligible contribution to the OTOC.
Only after $\kappa_{b}(t)$ has decayed and is irrelevant for the dynamics of the OTOC, finite size effects spoil the factorization.
The crossing contribution $\kappa_{(g)}(t)$ is suppressed at all times.

The relevance of the fourth cumulant $ \aveE{k_4(t)}_E$ for the OTOC dynamics, or more generally of higher cumulants of order $q$ for multi-time correlation functions, is a consequence of the correlations between matrix elements encoded in the full ETH ansatz~\eqref{ETHq} and their smooth frequency dependence, expressed in terms of $F_{e}^{(q)}(\vec \omega)$.
The latter is related to the free cumulants via Fourier transform, Eq.~\eqref{freekETH_Fourier}, which we study in the following.
More precisely, we define free cumulants $ \aveE{k_4(\vec \omega)}_E$ in the frequency domain by  replacing $A_{ij}(t)=A_{ij}e^{i\omega_{ij}t}$ with $A_{ij}(\omega)=A_{ij} g_\tau(\omega- \omega_{ij})$ in Eq.~\eqref{freekETH}.
Here, $g_\tau$ denotes a gaussian of width $\tau=0.02$, which approaches the delta distribution as $\tau \to 0$. 
The frequencies are subject to the constraint $\sum_{i=1}^q\omega_q=0$ following from the cyclic index structure.
The broadening of the delta distribution effectively corresponds to a regularized, i.e. finite time, Fourier transform in Eq.~\eqref{freekETH_Fourier}.
In a similar fashion, the ETH diagrams $\kappa_{(x)}$ can be translated into the frequency domain and their factorization properties can be studied there as well.

\begin{figure}[]
	\centering
	\includegraphics[width=8.5cm]{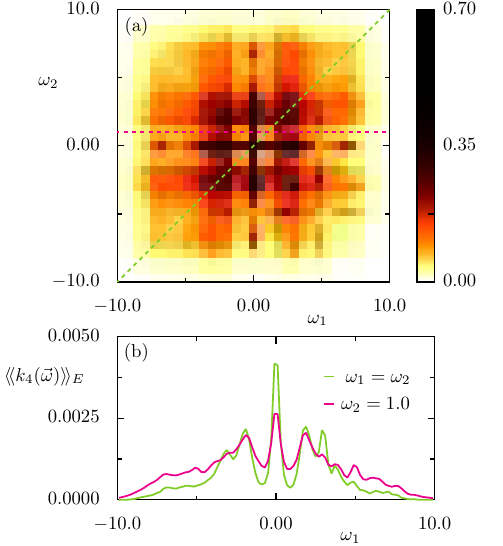}
	\caption{Frequency depencence of the fourth ETH cumulant $ \aveE{k_4(\vec \omega)}_E$ on the plane $\vec \omega = (\omega_1, \omega_2, -\omega_1, -\omega_2)$ for the extensive observable $\hat{A}$ with $N=16$ at microcanonical energy $E=0.25N$ and width $\Delta=1$ as (a) a function of both $\omega_1$ and $\omega_2$ and (b) along the lines $\omega_1=\omega_2$ and $\omega_2=1$ (indicated by dashed lines in (a)).}
	\label{fig:k4_frequency}
\end{figure}

In Fig.~\ref{fig:k4_frequency} we depict the fourth free cumulant $ \aveE{k_4(\vec \omega)}_E$ on the plane $\vec \omega = (\omega_1, \omega_2, -\omega_1, -\omega_2)$ as well as along two distinct lines.
As predicted by full ETH, the free cumulant, and hence the function $F_{e}^{(q)}(\vec \omega)$, smoothly depends on the the frequencies in a non-trivial way, i.e., it is non-zero and non-constant.
Similar as for the well known case of $F_{e}^{(2}(\omega)$ in standard ETH the function $F_{e}^{(q)}(\vec \omega)$ rapidly decays at large frequencies.

\begin{figure}[]
	\centering
	\includegraphics[width=8.5cm]{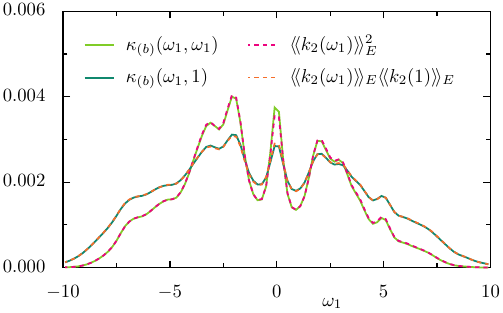}
	\caption{Factorization of $\kappa_{(b)}(\vec \omega)$ into free cumulants for for $\hat A$ and $N=16$ at microcanonical energy $E=0.25N$ and width $\Delta = 1$ along the lines indicated in Fig.~\ref{fig:k4_frequency}(a).}
	\label{fig:facto_cactus}
\end{figure}

Additionally, in Fig.~\ref{fig:facto_cactus} we show the factorization of $\kappa_{(b)}(\omega_1, \omega_2)= \aveE{k_4(\omega_1)}_E \aveE{k_4(\omega_2)}_E$ along the same lines in the frequency domain as in Fig.~\ref{fig:k4_frequency}(a). We again observe a smooth dependence on the frequencies and rapid decay towards large frequencies. This is the case for both $\kappa_{(b)}(\omega_1, \omega_2)$ and the factorized result. In fact both are almost indistinguishable on the shown scale except for small deviations at $\omega_1=0$.

This concludes the discussion of the OTOC, for which the above numerical experiments demonstrate, that the free probability approach to full ETH provides an accurate and simple description.

\subsection{Factorization of non-crossing ETH Diagrams}

The result above holds as a result of the validies of the \emph{Properties 1., 2.} discussed in Section \ref{sec_validity} above. We here establish their numerical validity.

\begin{figure}[t]
	\centering
	\includegraphics[width=8.5cm]{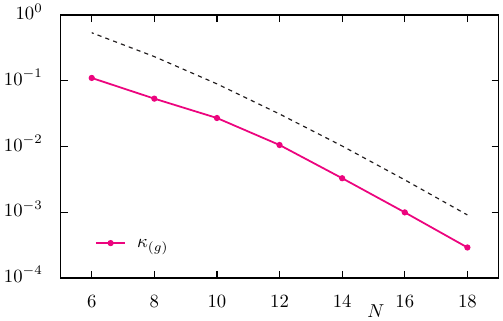}
	\caption{Suppression of the crossing diagram $\kappa_{(g)}(0)$ vs. system size $N$ for the observable $\hat{A}$ at microcanonical energy $E=0.25N$. The gray dashed line corresponds to the scaling $\sim e^{-S(E)}$.}
	\label{fig:crossing_factorization_systemsize}
\end{figure}

First of all, we illustrate \emph{Property 1}: the suppression of the crossing diagrams to scale as $Z_E^{-1}\sim e^{-S(E)}$, e.g. exponential suppression with system size $N$.
Indeed, we observe the predicted scaling, which we illustrate in Fig.~\ref{fig:crossing_factorization_systemsize}.

\begin{figure}[t]
	\centering
	\includegraphics[width=8.5cm]{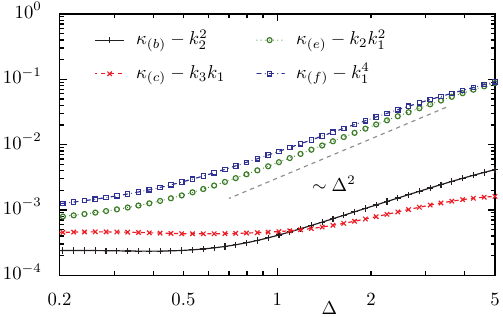}
	\caption{Factorization of ETH diagrams $\kappa_{(x)}(0)$ for the observable $A$ and system size $N=18$ at microcanonical energy $E=0.25N$ vs width $\Delta$ on a log-log scale. We abbreviate $k_q= \aveE{k_q(0)}_E$.}
	\label{fig:facto_diagrams}
\end{figure}

Secondly, we explore \emph{Property 2}: the factorization of the non-crossing ETH diagrams. These encode the contributions not entering the OTOC due to $ \aveE{k_1(A_0)}_E=0$.
We hence switch back to the extensive observable $A$ defined by Eq.~\eqref{eq:ext_observable} with $ \aveE{k_1(A)}_E\neq 0$ at the microcanonical energy $E=0.25 N$.
We begin by checking the factorization of the first class of diagrams in Eq.~\eqref{eq:facto_eth_diagrams} at equal time. (We abbreviate $\kappa_{(x)}= \kappa_{(x)}(0, 0, 0)$.) The diagrams $\kappa_{(b)}$, $\kappa_{(c)}$, $\kappa_{(e)}$, and $\kappa_{(f)}$ are expected to factorize as in Eq.~\eqref{eq:facto_eth_diagrams}, i.e., in terms of the diagrams corresponding to simple loops at the same energy. 
In particular, we study the dependence of the factorization on the microcanonical width $\Delta$.
For these diagrams, we show the difference between $\kappa_{(x)}$ and the corresponding factorized result in Fig.~\ref{fig:facto_diagrams} and observed the predicted scaling $\sim \Delta^2$ for not to small width $\Delta$. For smaller $\Delta$, similar as in Fig.~\ref{fig:diagonal_fluctuations}(c,d), we observe deviations from this scaling and an almost constant value, which can be attributed to the finite number of states in the microcanonical shell. 
This essentially confirms the predicted scaling for the first class of diagrams. 

\begin{figure}[h]
	\centering
	\includegraphics[width=8.5cm]{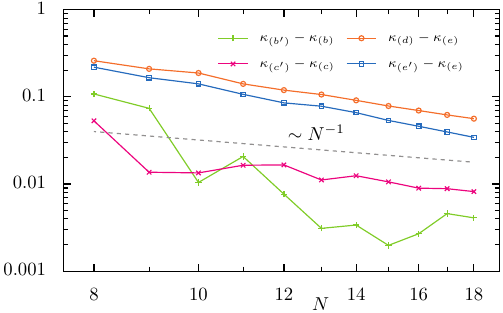} 
	\caption{Difference for ETH diagrams $(b)$ - $(e)$ vs.~system size $N$ for the observablae $A$ with microcanonical width $\Delta=1$ and microcanonical energy $E=0.25N$ on a log-log scale. 
    The dashed gray lines correspond to the scaling $\sim N^{-1}$.}
	\label{fig:eth_diagrams_vs_L}
\end{figure}

Finally, we consider the second class of diagrams, corresponding to simple loops on different energies.
These diagrams are given in \eqref{eq:facto_eth_diagrams2} and differ from the other class by $O(N^{-1})$, specifically
\begin{subequations}
\begin{align}
\kappa_{(b^\prime)}(0, 0, 0) & \simeq \kappa_{(b)}(0, 0, 0) + \mathcal O(N^{-1}) \ ,\\
\kappa_{(c^\prime)}(0, 0, 0) & \simeq \kappa_{(c)}(0, 0, 0) + \mathcal O(N^{-1}) \ ,\\
\kappa_{(e^\prime)}(0, 0, 0) & \simeq \kappa_{(e)}(0, 0, 0) + \mathcal O(N^{-1}) \ ,\\
\kappa_{(d)}(0, 0, 0) & \simeq \kappa_{(e)}(0, 0, 0) + \mathcal O(N^{-1}) \ .
\end{align}
\end{subequations} 
Thus in Fig.~\ref{fig:eth_diagrams_vs_L}, we depict the difference between the two classes as a function of system size $N$. 
We observe this difference to decay at least as $\sim N^{-1}$, indicating that both classes become equivalent in the thermodynamic limit and hence obey the same factorization.


\section{Conclusion and perspectives}
\label{sec:conclusion}

In this work, we extended the implications of the Eigenstate Thermalization Hypothesis (ETH) for multi-point correlation functions in the microcanonical ensemble, with a special emphasis on extensive observables. While canonical ensemble averages may deviate due to eigenstate-to-eigenstate and energy fluctuations, the microcanonical ensemble allows for these fluctuations to be minimized by controlling the width of the energy shell. This suppression of energy fluctuations, known as the standard  ETH, is particularly relevant for full ETH and multi-time correlation functions in nonintegrable systems. 

Our results highlight the significant role of free cumulants in decomposing multi-time correlation functions within the microcanonical ensemble. By establishing the connection between ETH and free probability, we have shown that the recursive nature of free cumulants provides a simplified yet powerful tool for characterizing complex correlations in a large class of physical observables. 
Our work highlights the importance of considering microcanonical ensembles in the study of quantum dynamics, offering a clearer and more precise characterization of the underlying statistical mechanics.

These findings open new avenues for future investigations. It would be interesting to understand how to extend this approach to the case of integrable systems \cite{essler2016quench, Vidmar2019, Goold2020}. Furthermore, these results may be applied to extend to higher-order the relation between ETH and hydrodynamics, where the overlap between diagonal and off-diagonal ETH plays an important role \cite{rakovszky2018diffusive, delacretaz2020heavy, capizzi2024hydrodynamics}. 

\begin{acknowledgments} 
FF acknowledges support by Deutsche Forschungsgemeinschaft (DFG), Project No. 453812159.
TP acknowledges support from ERC Advanced grant No.~101096208 -- QUEST, and Research Programme P1-0402 of Slovenian Research and
Innovation Agency (ARIS). SP acknowledges support from DFG under Germany’s
Excellence Strategy – Cluster of Excellence Matter and
Light for Quantum Computing (ML4Q) EXC 2004/1
– 390534769, and DFG Collaborative Research Center
(CRC) 183 Project No. 277101999 - project B02.
\end{acknowledgments}

\newpage

\begin{widetext}

\appendix

\section{Energy fluctuations in the canonical and microcanonical ensemble}
\label{app_Efluc}
We review the derivation of the energy fluctuations in the two ensembles.

\paragraph{Canonical ensemble}
The canonical ensemble is defined by $p_{E_\beta}(E_i)=e^{-\beta E_i}$ and $ Z_\beta = \Tr(e^{-\beta \hat H})$. First of all, the partition function reads
\begin{align}
   Z_\beta & = \sum_i {e^{-\beta E_i}} = \int dE_1 e^{S(E_1) - \beta E_1}
    = e^{S(E_\beta)-\beta E_\beta}\int dx e^{\frac {S''}2 x^2}
\end{align}
where, after substituting sums with integrals, one can compute this integral by saddle point in the limit $N\to\infty$. In this case, the saddle point is given by
$	\frac {\partial }{\partial E} (S(E) -\beta E)\Big |_{E^*} = 0$,
which identifies the thermal average $E^*=E_\beta = N e_\beta$, i.e. $S'(E^*)=\beta$. Then, we expand in the energy $E=E_\beta+ N x$ up to the second order. Using the convexity of the thermodynamic entropy, i.e. $S''=-|S''|$, the Gaussian integral leads to:
\begin{equation}
    Z_\beta = e^{S(E_\beta)-\beta E_\beta} \sqrt{2\pi} \, \frac 1{\sqrt{|S''|}}\ .
\end{equation}
In the same way, the energy fluctuations are evaluated from
\begin{align}
    \Delta_{E_\beta}^{(1)} & = \sum_i \frac{e^{-\beta E_i}}{Z_\beta} (E_i/N-e_\beta) \simeq  \frac {e^{S(E_\beta)-\beta E_\beta}}{Z_\beta} \frac 1N \int dx x e^{-\frac {|S''|}2 x^2} = 0
    \\
    \Delta_{E_\beta}^{(2)} & = \sum_i \frac{e^{-\beta E_i}}{Z_\beta} (E_i/N-e_\beta)^2 \simeq  \frac {e^{S(E_\beta)-\beta E_\beta}}{Z_\beta} \frac 1{N^2}\int dx x^2 e^{-\frac {|S''|}2 x^2} = \frac 1{N^2 |S''|} \ .
\end{align}

\paragraph{Microcanonical ensemble} We consider for definiteness a Gaussian microcanonical ensemble of width $\Delta$ as $p_{E, \Delta}= \exp ({-{(E_i-E)^2}/{2\Delta^2}})$.
The normalization for $N\to \infty$ is
\begin{align}
    \begin{split}
        \label{Ze_De_deriva}
       Z_E & = \sum_i p_{E}(E_i)  = \int dE_1 e^{S(E_1)-(E_1-E)^2/2\Delta^2} 
       \simeq e^{S(E)} \int dx e^{ S_E' x + \frac 12 \left ({S''}  - \frac 1{\Delta^2} \right )x^2 }
    \end{split}
\end{align}
where on the right-end side,  we solve the integral by a saddle point, expanding for $E_1=E+x$ for small $x$. Here $S_E'$ is the ``microcanonical temperature''. We denote
\begin{equation}    
    \label{Delta_app}
    \frac 1{\tilde \Delta^2} = \frac 1{\Delta^2} + |S''| \ .
\end{equation}
and solve the Gaussian integral in Eq.~\eqref{Ze_De_deriva}
\begin{align}
    \label{A4}
         Z_E  
       \simeq  e^{S(E)} \int dx e^{S' x -\frac{x^2}{2\tilde \Delta^2}} = e^{S(E)} \sqrt{2\pi} \tilde \Delta e^{\frac{\tilde \Delta^2 (S')^2}2}\ .
\end{align}
In the same way, the energy fluctuations are evaluated as
\begin{align}
    \Delta_{E_\beta}^{(1)} & = \sum_i \frac{e^{-\beta E_i}}{Z_\beta} (E_i/N-e_\beta) \simeq  \frac {e^{S(E_\beta)-\beta E_\beta}}{Z_\beta} \frac 1N \int dx x e^{S' x -\frac{x^2}{2\tilde \Delta^2}} = \frac {1}N \tilde \Delta^2 S'_E
    \\
    \Delta_{E_\beta}^{(2)} & = \sum_i \frac{e^{-\beta E_i}}{Z_\beta} (E_i/N-e_\beta)^2 \simeq  \frac {e^{S(E_\beta)-\beta E_\beta}}{Z_\beta} \frac 1{N^2}\int dx x^2 e^{S' x -\frac{x^2}{2\tilde \Delta^2}}
     = \frac {1}{N^2} \tilde \Delta^2 (1+S'^2 \tilde \Delta^2)\ .
\end{align}

At the leading order in $\tilde \Delta^2$ this leads to Eq.~\eqref{Delta_} of the main text.

\section{Derivation of Equation \eqref{freekETH_Fourier} }
\label{app_derive}
Let us review the derivation of Eq.~\eqref{freekETH_Fourier} from Eq.~\eqref{freekETH}. This is the microcanonical generalization of the calculation in Ref.\cite{pappalardi2022eigenstate}. Let us use time translational invariance and consider $ \aveE{k^{{\rm ETH}}_q(t_1, t_2, ... , t_{q-1}, t_q=0)}_E$ in Eq.~\eqref{freekETH} as given by
\begin{align}
     \aveE{k_q^{{\rm ETH}}(t_1, t_2, \dots t_{q-1}, 0)}_E & =  \sum_{i_1 \neq i_2 \dots \neq i_q }
    \delta_{\Delta}(E_i-E)
    e^{i t_1(E_{i_1}-E_{i_2}) +  t_2(E_{i_2}-E_{i_3}) + \dots  t_{q-1}(E_{i_{q}}-E_{i_{q-1}})} 
    {A_{i_1i_2}A_{i_2 i_3} \dots A_{i_{q}i_1}} 
    \\ & =
    \frac 1{Z_E} \sum_{i_1 \neq i_2 \dots \neq i_q }\frac{p_{E}(E_{i_1})}{Z_E} e^{i \vec t \cdot \vec \omega} e^{-(q-1) S(E^+)}  F^{(q)}_{e^+}(\vec \omega) \\
    & =
   \frac 1{Z_E}  \int dE_1 \dots dE_q p_E(E_{1}) e^{i \vec t \cdot \vec \omega} e^{S(E_1) + \dots S(E_q) - (q-1) S(E^+)}\, F^{(q)}_{e^+}(\vec \omega)
\end{align}
where from the first to the second line we have used $\vec 
\omega = (\omega_{i_1i_2}, \dots, \omega_{i_{q-1}i_q})$ with $\omega_{ij}=E_i-E_j$ and substituted the ETH ansatz [c.f. Eq.~\eqref{ETHq}] and from the second to the third we have exchanged the summation with the integral $\sum _{i_1}\to \int dE_1 e^{S(E_1)}$. 
We can thus Taylor expand the entropies around energy $E^+$ as
\begin{equation}
    S(E_i) = S(E^+ + (E_i-E^+)) = S(E^+) + S'(E^+)(E_i - E^+) + \frac 12 S''(E^+)(E_i - E^+)^2 + \dots \ .
\end{equation}
Then, by summing over all the energies one obtains
\begin{equation}
    \label{eq_expEntro}
    \sum_{i=1}^q S(E_i) = q S(E^+) +  S''(E^+) \sum_i (E_i - E^+)^2 + \dots \ ,
\end{equation} 
where the linear term in $E_i - E^+$ vanishes (due to $E^+ = (E_1 + E_2 + \dots E_q)/q$), while the quadratic term is subleading due to the thermodynamic property $S''(E^+) = - |S''(E^+)| \propto 1/N$. Since $E_i - E^+ \propto \vec \omega$ and $F_{e^+}(\vec \omega)$ is a smooth function that decays decays fast at large frequencies, we can neglect the second term in Eq.~\eqref{eq_expEntro}. The free cumulant then reads
\begin{align}
    \label{eq:22}
       \aveE{k_q^{{\rm ETH}}(\vec t)}_E = \frac 1{Z_E} \int dE_1 e^{-\frac{(E_1-E)^2}{2\Delta^2} + S(E^+)} \int dE_2 \dots dE_{E_q} e^{i \vec t \cdot \vec \omega} F^{(q)}_{e^+}(\vec \omega)   \ .
\end{align}
We can now rewrite 
\begin{align}
      E_1 & = E^+ + (E_1 - \frac{E_1 + E_2 + \dots E_q}q) = E^+ + \frac{q-1}{q}  (E_1 - E_2) + \frac{q-2}{q} (E_2-E_3) + \dots + \frac 1q (E_{q-1}-E_q) 
      \\ & = E^+ + \vec \ell_{q} \cdot \vec \omega \ ,
\end{align}
where we have defined the ladder vector 
\begin{equation}
    \label{eq:ladderQ}
    \vec \ell_q = \left ( \frac{q-1}{q}, \frac{q-2}{q} \dots , \frac{1}{q}\right ) \ .
\end{equation}
We substitute this into Eq.~\eqref{eq:22} and change integration variables $dE_1 dE_2 \dots dE_1 = dE^+ d\omega_1 d\omega_1 \dots d\omega_{q-1} $, leading to
\begin{align}
     \aveE{k_q^{{\rm ETH}}(\vec t)}_E
    & = \frac 1{Z_E} \int d \omega_1 \dots d\omega_{q-1} \int dE^+   e^{-\frac{(E^+-E)^2}{2 \Delta^2  } - \frac{E^+-E}{\Delta^2}  \vec \ell_q \cdot \vec \omega + S(E^+)}
    e^{i \vec t \cdot \vec \omega - \frac{(\vec \ell_q \cdot \vec \omega)^2}{2\Delta^2}} F^{(q)}_{e^+}(\vec \omega) 
    \\
    & = \frac {e^{S(E)}}{Z_E} \int d \omega_1 \dots d\omega_{q-1} e^{i \vec t \cdot \vec \omega - \frac{(\vec \ell_q \cdot \vec \omega)^2}{2\Delta^2}} \int dx   e^{-\frac{x^2}{2 \tilde \Delta^2  } - \frac{x}{\Delta^2}  \vec \ell_q \cdot \vec \omega + S'x}
    \left [ F^{(q)}_{e}(\vec \omega) +x[F^{(q)}_{e}(\vec \omega)]' + \frac {x^2}2 [F^{(q)}_{e}(\vec \omega)]''
    \right ]
    \\
    & = \int d \omega_1 \dots d\omega_{q-1} e^{i \vec t \cdot \vec \omega}  
    F^{(q)}_{e}(\vec \omega) e^{-S'(\vec \ell_q\cdot \vec \omega)}
    + \mathcal O(\Delta^2)
\end{align}
As done in the calculations above, in the first line, to solve the integral over $E^+$ by saddle point, we expand  around $E^+=E+x$ and we use the definition of $\tilde \Delta$ in Eq.~\eqref{Delta_app}. We then solve the gaussian integral and simplify the quadratic terms in $\vec \omega^2$ using the small $\tilde \Delta = \Delta$ limit. 
 This corresponds to the desired Eq.~\eqref{freekETH_Fourier}.

\section{Factorization of ETH Diagrams}
\label{app:factorization}

Here, we provide additional examples on how to derive the factorization of the ETH diagrams stated in Eq.~\eqref{eq:facto_eth_diagrams}. 
For instance for the diagram $(c)$ one has 
    \begin{align}
        \begin{split}
        \kappa_{(c)}(t_1, t_2, t_3) &  = \sum_{i} \frac{p_{E}(E_i)}{Z_E} A(t_1)_{ii} \sum_{m\neq k: \neq i}
        A(t_2)_{ik}A(t_3)_{km} A_{mi}
        \\ & =  \aveE{k_1^{  i} k_3^{  i}(t_2, t_3)}_E 
    =  \aveE{k_1}_E\;  \aveE{k_3(t_2, t_3, 0)}_E + \mathcal O(\Delta^2)            
        \end{split}
    \end{align}
while the factorization of diagram $(f)$ follows via
\begin{align}
    \kappa_{(f)}(t_1, t_2, t_3) &  =  \frac{1}{Z_E} \sum_{i}p_{E}(E_i) [k^{i}_1]^4  =  \aveE{[k^{i}_1]^4 }_E = 
    \aveE{k^{{\rm ETH}}_1}^4_E + \mathcal O(\Delta^2) \ .
\end{align}
The factorization of diagram $(d)$, which belongs to the second class of diagrams factorizes up to an additional $\mathcal{O}(N^{-1})$ correction, as can be seen from 
\begin{align}
    \begin{split}
            \kappa_{(d)}(t_1, t_2, t_3) &  = \sum_{i} \frac{p_{E}(E_i)}{Z_E}  A_{ii} 
       \sum_{j:j\neq i} A(t_1)_{ij}A(t_3)_{ji} A_{jj}  
       \\ & = \sum_{i} \frac{p_{E}(E_i)}{Z_E}  k_1^{i} 
       \sum_{j:j\neq i} A(t_1)_{ij}A(t_3)_{ji}  k_1^{j}
         \\ & = \sum_{i} \frac{p_{E}(E_i)}{Z_E}  k_1^{i} 
       \sum_{j:j\neq i} A(t_1)_{ij}A(t_3)_{ji} \left ( k_1^{i}  + \frac{\partial k_1^{i}}{\partial e_i} \frac{\omega_{ji}}N + \frac 12 \frac{\partial^2 k_1^{i}}{ \partial e_i^2} \frac{\omega_{ji}^2}N^2 \right )
        \\ & =  \aveE{ k_1^{i}  k_2^{i}(t_1, t_3)  k_1^{i}}_E + \mathcal O( N^{-1}) 
        =  \aveE{ k_1^{{\rm ETH}}}^2_E  \aveE{ k_2^{{\rm ETH}}(t_1, t_3)}_E + \mathcal O(\Delta^2, N^{-1}) \ . 
    \end{split}
\end{align}
As outlined in the main text for the factorization of the diagram $(b^\prime)$ in Eq.~\eqref{kb1exp_}, the frequency dependent term $\sim \omega_{ji}$ becomes negligible at large $N$ as it is multiplied by $A(t_1)_{ij}A(t_3)_{ji}$, which decays rapidly at large frequency.
    
\end{widetext}

\bibliography{biblio}

\end{document}